\documentclass[12pt]{article}
\usepackage{amsfonts}
\usepackage{amssymb}
\usepackage{amsmath}
\usepackage{amsthm}
\renewcommand{\div}{\mathop{\rm div}\nolimits}

\makeatletter
\renewcommand{\section}{\@startsection{section}{1}{0pt}%
{3.5ex plus 1ex minus .2ex}{2.3ex plus .2ex}%
{\large\bf}}
\makeatother
\newtheorem{theorem}{Theorem}
\newtheorem{lemma}[theorem]{Lemma}
\author{Alexey V. Golovnev  \\
{  }\\
{\small \it Saint-Petersburg State University}\\
{\small\it high energy physics department,}\\
{\small\it Ulianovskaya ul., 1;}\\
{\small\it Stary Petergof, Saint-Petersburg, Russia}\\
{\small agolovnev@yandex.ru}\\
{\small alex@amber.ff.phys.spbu.ru}\\
{\small +7(812)4287339}}
\title{DIRAC QUANTIZATION\\ OF FREE MOTION\\ ON CURVED SURFACES}
\date{ }
\begin{document}
\maketitle
\abstract

We give an explicit operator realization of Dirac quantization of free particle motion
on a surface of codimension 1. It is shown that the Dirac recipe is ambiguous and
a natural way of fixing this problem is proposed. We also introduce a modification
of Dirac procedure which yields zero quantum potential. Some problems of abelian conversion 
quantization are pointed out.

\vspace{2ex}


\vspace{2ex}

{\bf keywords}: canonical quantization, Dirac brackets, quantum potential.

\newpage
\section{Introduction}
We consider a problem of quantum motion in curved spaces. It is well-known that in the case of Euclidean
spaces the correct Hamiltonian is ${\hat H}=-\frac{{\hbar}^2}{2}\Delta$. Podolsky \cite{Podolsky} in 1928 proposed that
for arbitrary space it should be replaced by ${\hat H}=-\frac{{\hbar}^2}{2}{\Delta}_{LB}$ with ${\Delta}_{LB}$
being the Laplace-Beltrami operator. This postulate is a direct and geometrically clear generalization
of the dynamics in Euclidean spaces.

If one wants to get the theory by some canonical procedure, he or she encounters a severe problem. For
any given classical theory there is an infinite number of quantum theories with a proper $\hbar\to 0$ limit.
Quantization is not unique. In Euclidean spaces Dirac recipe in Cartesian coordinates yields
experimentally correct result. In curved spaces we do not have a notion of Cartesian coordinates and can't
make a choice of the theory. A possible way out is to embed the space under consideration into some Euclidean space
and to quantize the new theory as a theory with second-class constraints by Dirac brackets formalism \cite{Dirac} or
by abelian conversion method \cite{Faddeev,Batalin}. The results are different and depend on the embedding. In this article
we restrict ourselves to codimension 1 surfaces.

In section 2 we describe the Dirac approach to motion on spheres \cite{KleSha,KlaSha} and develop an explicit operator
realization of it. If one demands the momenta to be differentials (instead of self-adjointness condition) the
quantum potential would be zero. In section 3 we generalize our consideration to the case of arbitrary surface
and show that the Dirac procedure is ambiguous. Dirac quantum potential depends on the choice of equation
of surface; a way of fixing this freedom is proposed. At the same time the zero-potential quantization as in
section 2\ is well defined. The zero potential may be obtained for spheres by the abelian conversion method \cite{KleSha} too. In section 4 we point
out some obstructions on the way of generalizing it to arbitrary surfaces and show that in general case
one can't get zero potential by abelian conversion.

\section {Dirac quantization for spheres}
We start with a free particle motion on $(n-1)$-dimensional sphere, $\sum\limits_{i=1}^{n}x^{2}_{i}=R^2$, in $n$-dimensional
Euclidean space. It can be considered as a system with two second-class constraints \cite{KleSha}
\begin{equation}
\label{1const}
\phi_1\equiv\sum_{i=1}^{n}x_i^2-R^2=0,
\end{equation}
\begin{equation}
\label{2const}
\phi_2\equiv\sum_{i=1}^{n}x_ip_i=0
\end{equation}
where $p_i$ are canonical momenta. The Poisson bracket  $\{\phi_1,\phi_2\}=2{\overrightarrow x}^2$ is not zero
(it is the definition of second-class constraints), and hence in quantum theory the constraints (\ref{1const}) and
(\ref{2const}) can not be set equal zero simultaneously
even for a physical sector \cite{Dirac}. This problem can be overcome by introducing the Dirac 
brackets:
\begin{equation}
\label{Dirac}
\{ f,g\}_{\cal D}=\{ f,g\}-\sum_{a=1}^2\sum_{b=1}^2\{ f,{\phi}_a\}\Delta_{ab}\{ \phi_b,g\},
\end{equation}
where $\Delta_{ab}$ is the matrix inverse of $\{ \phi_a,\phi_b\}$. Now $\{\phi_1,\phi_2\}_D=0$ and for canonical variables
we have \cite{KleSha}
\begin{equation}
\label{sph1}
\{ x_i,x_j\}_{\cal D}=0,
\end{equation}
\begin{equation}
\label{sph2}
\{ x_i,p_j\}_{\cal D}=\delta_{ij}-\frac{x_ix_j}{{\overrightarrow x}^2},
\end{equation}
\begin{equation}
\label{sph3}
\{ p_i,p_j\}_{\cal D}=\frac{1}{{\overrightarrow x}^2}(p_ix_j-p_jx_i).
\end{equation}

Dirac bracket is degenerate and does not define any symplectic manifold but it can be regarded as a Poisson
structure \cite{Reyman} obtained by factorization of original Poisson bracket algebra over motions in unphysical direction.
One can get it by the following replacement:
\begin{equation}
\label{factor}
{\overrightarrow P} \rightarrow {\overrightarrow P} - 
\left (\frac{\overrightarrow x}{|\overrightarrow x|} \cdot {\overrightarrow P}\right )
\frac{\overrightarrow x}{|\overrightarrow x|},
\end{equation}
so that all different values of radial momentum are identified. Another possible interpretation is made in \cite{Brat1,Brat2} in terms of
first-class functions algebra factorized over functions vanishing on the constraint surface.

Once we have the Dirac structure, the quantization can be performed in the usual way \cite{Dirac}. From 
(\ref{sph1})-(\ref{sph3}) we get
\begin{equation}
\label{sphop1}
[\hat x_i,\hat x_j]=0,
\end{equation}
\begin{equation}
\label{sphop2}
[\hat x_i,\hat p_j]=i\hbar\left (\delta_{ij}\hat I-\frac{\hat x_i\hat x_j}{\sum\limits_{l=1}^n\hat x_l^2}\right ),
\end{equation}
\begin{equation}
\label{sphop3}
[\hat p_i,\hat p_j]=\frac{i\hbar}{\sum\limits_{l=1}^n\hat x_l^2}\left (\hat p_i\hat x_j-\hat p_j\hat x_i\right ).
\end{equation}
In (\ref{sphop3}) the operator ordering problem is solved; we show that this ordering is correct 
(satisfies the Jacobi identity) by providing
an explicit operator realization of the algebra (\ref{sphop1})-(\ref{sphop3}). We choose coordinate operators to be
the usual ones $\hat x_i=x_i\hat I$ and search for corresponding 
differential operators of momenta. One could solve the task by use
of (\ref{factor}):
$$-i\hbar{\overrightarrow\bigtriangledown} \rightarrow -i\hbar{\overrightarrow\bigtriangledown} - 
\frac{\overrightarrow x}{|\overrightarrow x|}
\left (\frac{\overrightarrow x}{|\overrightarrow x|} \cdot \left (-i\hbar{\overrightarrow\bigtriangledown}\right )\right )
.$$ 
This choice is in some sense unique. Indeed, we demand $\hat p_i$ to be differentiations, i.e. to
obey the Leibnitz rule. From (\ref{sphop2}) one gets 
${\hat p}_i(x_j)=-i\hbar\left (\delta_{ij}-\frac{x_ix_j}{{\overrightarrow x}^2}\right )$. By the Leibnitz rule we have for
any polynomial
\begin{multline*}
{\hat p}_k\left (\sum_{\{\alpha\}}C_{\{\alpha\}}\prod_{i=1}^{|\alpha|}x_{\alpha_i}\right )=-i\hbar\sum_{\{\alpha\}}C_{\{\alpha\}}\cdot\\
\cdot\left (\sum_{i=1}^{|\alpha|}\delta_{k\alpha_i}\left (\prod_{j=1}^{i-1}x_{\alpha_j}\right )\left (\prod_{j=i+1}^{|\alpha|}x_{\alpha_j}\right )
-|\alpha|\frac{x_k\prod\limits_{i=1}^{|\alpha|}x_{\alpha_i}}{\sum\limits_{l=1}^nx_l^2}\right )
\end{multline*}
and extend this definition to analytic functions by continuity:
\begin{equation}
\label{sphdmom}
{\hat p}_i=-i\hbar\left (\frac{\partial}{\partial x_i}-\frac{x_i}{|{\overrightarrow x}|}\sum_{j=1}^n\frac{x_j}{|{\overrightarrow x}|}
\frac{\partial}{\partial x_j}\right );
\end{equation}
it's the projection (\ref{factor}) of basis vectors $-i\hbar\frac{\partial}{\partial x_i}$ onto the surface (\ref{1const}).

The calculation of commutators in (\ref{sphop3}) is straight-forward. It yields
\begin{multline*}
[{\hat p}_i,{\hat p}_j]=-\frac{\hbar^2}{{\overrightarrow x}^2}\left (x_i\frac{\partial}{\partial x_j}-x_j\frac{\partial}{\partial x_i}\right )=\\
=\frac{i\hbar}{{\overrightarrow x}^2}(x_j{\hat p}_i-x_i{\hat p}_j)=\frac{i\hbar}{{\overrightarrow x}^2}({\hat p}_ix_j-{\hat p}_jx_i).
\end{multline*}
So, the second constraint is satisfied identically,
$\sum\limits_{i=1}^n{\hat x}_i{\hat p}_i\equiv 0$, and we fix the physical sector simply by
$\Psi_{phys}=\psi (x)\delta\left (\sum\limits_{i=1}^nx_i^2-R^2\right )$.

The problem is that $\hat p_i$ are not self-adjoint. At the sacrifice of Leibnitz rule we can introduce  new self-adjoint momenta:
\begin{multline*}
{\hat{\tilde p}}_i=\frac{1}{2}({\hat p}_i+{\hat p}_i^{\dag})=\\=-i\hbar
\left (\frac{\partial}{\partial x_i}-\frac{x_i}{|{\overrightarrow x}|}\sum_{j=1}^n\frac{x_j}{|{\overrightarrow x}|}\frac{\partial}{\partial x_j}
-\frac{1}{2}\sum_{j=1}^{n}\frac{\partial}{\partial x_j}\left (\frac{x_ix_j}{{\overrightarrow x}^2}\right ){\hat I}\right )=\\
={\hat p}_i+i\hbar\frac{1}{2}\sum_{j=1}^{n}\frac{\partial}{\partial x_j}\left (\frac{x_ix_j}{{\overrightarrow x}^2}\right ){\hat I}=
{\hat p}_i+i\hbar\frac{n-1}{2}\cdot\frac{x_i}{{\overrightarrow x}^2}{\hat I}.
\end{multline*}
It is easy to check that the algebra (\ref{sphop1})-(\ref{sphop3}) remains the same. The constraint (\ref{2const}) is now
${\hat \phi}_2=\sum\limits_{i=1}^n({\hat x}_i{\hat{\tilde p}}_i+({\hat x}_i{\hat{\tilde p}}_i)^{\dag})\equiv 0$.

\begin{theorem}
The Hamiltonian ${\hat H}^{\cal(D)}=\frac{1}{2}\sum\limits_{i=1}^n{\hat{\tilde p}}_i^2$ contains
quantum potential $V_q^{\cal(D)}=\frac{\hbar^2(n-1)^2}{8R^2}$.
\end{theorem}
\begin{proof}
By direct calculation we have
$${\hat{\tilde p}}_i\left (\psi (x)\cdot\delta\left (\sum\limits_{i=1}^nx_i^2-R^2\right )\right )=
{\hat{\tilde p}}_i(\psi (x))\cdot\delta\left (\sum\limits_{i=1}^nx_i^2-R^2\right ),$$
\begin{multline*}
{\hat H}^{\cal(D)}\left (\psi (x)\right )=\\=
\left(\frac12\sum\limits_{i=1}^n\left({\hat p_i}^2+i\hbar(n-1)\frac{x_i}{|\overrightarrow x|}{\hat p_i}\right)
\right)\Psi(x)
+\frac{\hbar^2(n-1)^2}{8R^2}\Psi(x)=\\=
-\frac{\hbar^2}{2}\left(\sum\limits_{i=1}^n\frac{\partial^2}{\partial x_i^2}-
\sum\limits_{i=1}^n\frac{x_i}{|\overrightarrow x|}\frac{\partial}{\partial x_i}\cdot
\sum\limits_{j=1}^n\frac{x_j}{|\overrightarrow x|}\frac{\partial}{\partial x_j}-\right.\\\left.-
\frac{n-1}{|\overrightarrow x|}\sum\limits_{i=1}^n\frac{x_i}{|\overrightarrow x|}\frac{\partial}{\partial x_i}
\right)\Psi(x)+\frac{\hbar^2(n-1)^2}{8R^2}\Psi(x)
=\\=
-\frac{\hbar^2}{2}\Delta_{LB}(\psi(x))+\frac{\hbar^2(n-1)^2}{8R^2}\psi(x),
\end{multline*}
where $\Delta_{LB}$ is the Laplace-Beltrami operator on sphere (for calculation of $\Delta_{LB}$ see section 3).
So, the Dirac quantum potential is $V_q^{\cal(D)}=\frac{\hbar^2(n-1)^2}{8R^2}$.
\end{proof}

This result coincides with the conclusion of \cite{KleSha}, but the approach of
\cite{KleSha} is purely algebraic. What we presented here is an explicit operator realization of it,
which clarifies the geometric properties.

The same procedure may lead to Podolsky theory if one takes the definition (\ref{sphdmom}) and
Hamiltonian ${\hat H}^{\cal(P)}=\frac{1}{2}\sum\limits_{i=1}^n{\hat p}^{\dag}_ip_i$ which equals 
$-\frac{\hbar^2}{2}\Delta_{LB}$ for the physical sector functions. The quantum potential is zero:
$V_q^{\cal(P)}=0$. Thus one preserves an important property of momenta operators, the Leibnitz rule, so that they are
differentials on the algebra of smooth functions. These operators are not self-adjoint and can't represent observables.
But they do not have any clear physical meaning being projections of generators of motions along the coordinate lines of 
n-dimensional flat space, which are somewhat esoteric for an observer living on the sphere.
Natural observables on the sphere are generators of $SO(n)$ rotations, and they are self-adjoint
(proportional to $i[{\hat p}_i,{\hat p}_j]$).

We should note that operators  ${\hat p}_i$ are not self-adjoint with respect to Lebesgue measure
in the other space ${\mathbb R}^n$. In order to get zero quantum potential with self-adjoint momenta
one could try to find another measure for which these operators would be self-adjoint. But the potential
$\frac{\hbar^2(n-1)^2}{8R^2}$ can be obtained algebraically [10], without use of any particular measure.
Moreover, the desired measure does not exist. Indeed, for a measure $G(x)d^nx$ the operators
 ${\hat p}_i$ would be symmetric if and only if $\left (\frac{\partial}{\partial x_i}-\frac{x_i}{R}
\sum\limits_{j=1}^n\frac{x_j}{R}
\frac{\partial}{\partial x_j}\right )G(x)=\frac{(n-1)x_i}{R^2}G(x)$. After multiplication by $x_i$ and
summation over $i$ one has $G=0$. Due to the reasons mentioned in the Introduction we 
prefer to quantize in Cartesian coordinates with the standard Lebesgue measure.

\section{Dirac quantization for arbitrary surfaces}
We consider motions on a codimension 1 surface $f(x_i)=0$. This theory has two constraints \cite{KlaSha}
\begin{equation}
\label{arbcon1}
\phi_1\equiv f(x)=0,
\end{equation}
\begin{equation}
\label{arbcon2}
\phi_2\equiv\sum_{i=1}^n(\partial_if)p_i=0.
\end{equation}
These constraints are of the second class because $\{\phi_1,\phi_2\}=({\overrightarrow \bigtriangledown}f)^2\ne0$. We
introduce the Dirac brackets by (\ref{Dirac}) and get
\begin{equation}
\label{arb1}
\{ x_i,x_j\}_{\cal D}=0,
\end{equation}
\begin{equation}
\label{arb2}
\{ x_i,p_j\}_{\cal D}=\delta_{ij}-\frac{(\partial_if)(\partial_jf)}{({\overrightarrow \bigtriangledown}f)^2},
\end{equation}
\begin{equation}
\label{arb3}
\{ p_i,p_j\}_{\cal D}=\frac{1}{({\overrightarrow \bigtriangledown}f)^2}\sum_{k=1}^{n}((\partial_jf)(\partial^2_{ik}f)-
(\partial_if)(\partial^2_{jk}f))p_k.
\end{equation}

We propose the following operators for the quantum description
with non-selfadjoint momenta: ${\hat x}_i=x_i{\hat I}$ and
$${\hat p}_i=-i\hbar\left (\frac{\partial}{\partial x_i}-\frac{(\partial_if)}{|{\overrightarrow\bigtriangledown}f|}
\sum\limits_{j=1}^n\frac{(\partial_jf)}{|{\overrightarrow\bigtriangledown}f|}\frac{\partial}{\partial x_j}\right ).$$
Here we used the factorization over unphysical motions again. 
\begin{lemma}
The commutator algebra (corresponding to
(\ref{arb1})-(\ref{arb3})) is
\begin{equation}
\label{arbop1}
[{\hat x}_i,{\hat x}_j]=0,
\end{equation}
\begin{equation}
\label{arbop2}
[{\hat x}_i,{\hat p}_j]=i\hbar\left (\delta_{ij}-\frac{(\partial_if)(\partial_jf)}{({\overrightarrow \bigtriangledown}f)^2}\right ){\hat I},
\end{equation}
\begin{equation}
\label{arbop3}
[{\hat p}_i,{\hat p}_j]=\frac{i\hbar}{({\overrightarrow \bigtriangledown}f)^2}\sum_{k=1}^{n}((\partial_jf)(\partial^2_{ik}f)-
(\partial_if)(\partial^2_{jk}f)){\hat p}_k.
\end{equation}
\end{lemma}
\begin{proof}
For operators under consideration  (\ref{arbop1}) and (\ref{arbop2}) are obvious while (\ref{arbop3}) can be proved by a
direct calculation:
\begin{multline*}
[{\hat p}_i,{\hat p}_j]=-\hbar^2\left(\left[\frac{\partial}{\partial x_i}\ ,\ 
-\frac{(\partial_j f)}{|{\overrightarrow\bigtriangledown}f|}
\sum\limits_{k=1}^n\frac{(\partial_k f)}{|{\overrightarrow\bigtriangledown}f|}\frac{\partial}{\partial x_k}\right]\right.+\\
+\left.\left[-\frac{(\partial_i f)}{|{\overrightarrow\bigtriangledown}f|}
\sum\limits_{m=1}^n\frac{(\partial_m f)}{|{\overrightarrow\bigtriangledown}f|}\frac{\partial}{\partial x_m}\ ,\ 
\frac{\partial}{\partial x_j}\right]+\right.\\+\left.
\left[\frac{(\partial_i f)}{|{\overrightarrow\bigtriangledown}f|}
\sum\limits_{m=1}^n\frac{(\partial_m f)}{|{\overrightarrow\bigtriangledown}f|}\frac{\partial}{\partial x_m}\ ,\ 
\frac{(\partial_j f)}{|{\overrightarrow\bigtriangledown}f|}
\sum\limits_{k=1}^n\frac{(\partial_k f)}{|{\overrightarrow\bigtriangledown}f|}\frac{\partial}{\partial x_k}\right]\right)=\\=
\hbar^2\sum_{k=1}^{n}\frac{(\partial_jf)(\partial^2_{ik}f)-
(\partial_if)(\partial^2_{jk}f)}{({\overrightarrow \bigtriangledown}f)^2}\cdot\\ \cdot
\left(\frac{\partial}{\partial x_k}-\frac{(\partial_k f)}{|{\overrightarrow\bigtriangledown}f|}
\sum\limits_{m=1}^n\frac{(\partial_m f)}{|{\overrightarrow\bigtriangledown}f|}\frac{\partial}{\partial x_m}\right ).
\end{multline*}
\end{proof}
One also has $\sum\limits_{i=1}^n(\partial_if){\hat p}_i\equiv 0$. The physical sector is defined by
$\Psi_{phys}=\psi(x)\delta(f(x))$ and 
the Hamiltonian ${\hat H}^{\cal(P)}=\frac{1}{2}\sum\limits_{i=1}^n{\hat p}_i^{\dag}{\hat p}_i$ equals
\begin{multline*}
{\hat H}^{\cal(P)}=\frac{1}{2}\sum_{i=1}^n\left ( {\hat p}_i+i\hbar
\left (\sum_{j=1}^n\frac{\partial}{\partial x_j}\frac{(\partial_if)(\partial_jf)}{({\overrightarrow \bigtriangledown}f)^2}\right){\hat I}\right){\hat p}_i=\\
=-\frac{\hbar^2}{2}\left (\sum_{i=1}^{n}\frac{\partial^2}{{\partial x_i}^2}-\sum_{i=1}^{n}\frac{(\partial_if)}{|{\overrightarrow\bigtriangledown}f|}
\frac{\partial}{\partial x_i}\cdot\sum_{j=1}^{n}\frac{(\partial_jf)}{|{\overrightarrow\bigtriangledown}f|}
\frac{\partial}{\partial x_j}-\right.\\
-\left.\left(\sum_{i=1}^{n}\frac{\partial}{\partial x_i}\frac{(\partial_if)}{|{\overrightarrow\bigtriangledown}f|}\right)
\sum_{j=1}^{n}\frac{(\partial_jf)}{|{\overrightarrow\bigtriangledown}f|}
\frac{\partial}{\partial x_j}\right)=\\
=-\frac{\hbar^2}{2}\left({\tilde\Delta}-\left(\frac{\partial}{\partial\overrightarrow n}\right)^2-\div({\overrightarrow n})\cdot\frac{\partial}{\partial\overrightarrow n}\right)
\end{multline*}
where $\tilde\Delta$ is the Laplace operator in the Euclidean space and ${\overrightarrow n}=\frac{\overrightarrow\bigtriangledown f}
{|{\overrightarrow\bigtriangledown f}|}$ is a unit vector normal to the surface (\ref{arbcon1}). 

Now we follow the standard Dirac procedure and replace our operators by self-adjoint ones:
\begin{equation}
\label{new}
{\hat{\tilde p}}_i={\hat p}_i+
\frac{i\hbar}{2}\sum_{j=1}^n\left(\frac{\partial}{\partial x_j}\left(
\frac{(\partial_if)(\partial_jf)}{({\overrightarrow \bigtriangledown}f)^2}\right)\right).
\end{equation}
It violates the relation (\ref{arbop3}), but one can overcome this problem by changing the operator ordering. 
Indeed, from (\ref{arbop3}) we have
$$[{\hat p}_i^{\dag},{\hat p}_j^{\dag}]=i\hbar\sum_{k=1}^n{\hat p}_k^{\dag}
\frac{(\partial_jf)(\partial^2_{ik}f)-
(\partial_if)(\partial^2_{jk}f)}{({\overrightarrow \bigtriangledown}f)^2}$$
for ${\hat p}_i^{\dag}={\hat p}_i+
i\hbar\sum\limits_{j=1}^n\left(\frac{\partial}{\partial x_j}\left(
\frac{(\partial_if)(\partial_jf)}{({\overrightarrow \bigtriangledown}f)^2}\right)\right)$. It is not difficult to
deduce the following commutational relation from it:
\begin{multline*}
[{\hat{\tilde p}}_i,{\hat{\tilde p}}_j]=\frac{i\hbar}{2}\sum_{k=1}^n
\left (\frac{(\partial_jf)(\partial^2_{ik}f)-
(\partial_if)(\partial^2_{jk}f)}{({\overrightarrow \bigtriangledown}f)^2}{\hat{\tilde p}}_k\right.+\\
+\left.{\hat{\tilde p}}_k
\frac{(\partial_jf)(\partial^2_{ik}f)-
(\partial_if)(\partial^2_{jk}f)}{({\overrightarrow \bigtriangledown}f)^2}\right),
\end{multline*}
which differs from (\ref{arbop3}) only by operator ordering. The Hamiltonian 
${\hat H}^{\cal (D)}=\frac{1}{2}\sum\limits_{i=1}^n{\hat{\tilde p}}_i^2={\hat H}^{\cal (P)}+V_q(x)$
contains the quantum potential
\begin{multline*}
V_q=
-\frac{\hbar^2}{8}\sum_{i=1}^n\left(\sum_{j=1}^n\frac{\partial}{\partial x_j}
\frac{(\partial_if)(\partial_jf)}{({\overrightarrow \bigtriangledown}f)^2}\right)^2+\\
+\frac{\hbar^2}{4}\sum_{i=1}^n\left(\frac{\partial}{\partial x_i}-
\sum_{k=1}^n
\frac{(\partial_if)(\partial_kf)}{({\overrightarrow \bigtriangledown}f)^2}
\frac{\partial}{\partial x_k}\right)
\left(\sum_{j=1}^n\frac{\partial}{\partial x_j}\frac{(\partial_jf)(\partial_if)}{({\overrightarrow \bigtriangledown}f)^2}\right).
\end{multline*}
Unfortunately both Hamiltonians, ${\hat H}^{\cal (D)}$ and ${\hat H}^{\cal (P)}$, are ambiguous; they take
different values for those functions which represent one and the same surface.
The problem exists even for spheres. We first prove it for ${\hat H}^{\cal (D)}$.

\begin{theorem}
Dirac quantization procedure is ambiguous.
\end{theorem}
\begin{proof}
Indeed, any surface can be represented by its tangent pa\-ra\-bo\-loid at some point:
$f(y)=y_n-\frac{1}{2}\sum\limits_{\alpha=1}^{n-1}k_{\alpha}y_{\alpha}^2+{\cal O}(y_{\alpha}^3)$,\quad 
$y_{\alpha}$'s are Cartesian coordinates. We have
$n_n=-\frac{\partial_nf}{|\overrightarrow\bigtriangledown f|}=-(1+\sum\limits_{\alpha=1}^{n-1}k_{\alpha}^2y_{\alpha}^2)^{-1/2}
+{\cal O}(y_{\alpha}^2)=-1+{\cal O}(y_{\alpha}^2)$;  $n_{\alpha}=-\frac{\partial_{\alpha}f}{|\overrightarrow\bigtriangledown f|}=
k_{\alpha}y_{\alpha}+{\cal O}(y_{\alpha}^2)$ and $\partial^2_{in}f=0$.
One can neglect ${\cal O}(y_{\alpha}^3)$ terms in the calculation, because
$\sum\limits_{j=1}^n\frac{\partial}{\partial y_j}\frac{(\partial_jf)(\partial_nf)}{({\overrightarrow \bigtriangledown}f)^2}=
-\sum\limits_{\alpha=1}^{n-1}k_{\alpha}+{\cal O}(y_{\alpha})$;
$\sum\limits_{j=1}^n\frac{\partial}{\partial y_j}\frac{(\partial_jf)(\partial_{\alpha}f)}{({\overrightarrow \bigtriangledown}f)^2}
=\sum\limits_{\beta=1}^{n-1}k_{\alpha}k_{\beta}y_{\alpha}+k_{\alpha}^2y_{\alpha}+{\cal O}(y_{\alpha}^2)$ and
$\sum\limits_{k=1}^n\sum\limits_{j=1}^n\
\frac{(\partial_if)(\partial_kf)}{({\overrightarrow \bigtriangledown}f)^2}
\frac{\partial}{\partial y_k}
\frac{\partial}{\partial y_j}\frac{(\partial_jf)(\partial_if)}{({\overrightarrow \bigtriangledown}f)^2}
={\cal O}(y_{\alpha})$ (if $k=n$ then we get zero identically, if $k\neq n$ then
$\partial_kf={\cal O}(y_{\alpha})$).
The obtained quantum potential in the vicinity of the point $\overrightarrow y=0$ equals
$$V_q=\frac{\hbar^2}{8}\left(\left(\sum\limits_{\alpha=1}^{n-1}k_{\alpha}\right)^2+2\sum\limits_{\alpha=1}^{n-1}k_{\alpha}^2\right)+{\cal O}(y_{\alpha}).$$
For a sphere the principal curvatures are $k_{\alpha}=\frac{1}{R}$ and at the chosen point we have $V_q=\frac{\hbar(n^2-1)}{8R^2}$ which differs
from the result of \cite{KleSha} and section 2. So, the Dirac recipe is ambiguous.
\end{proof}

To fix the freedom, let's consider a curvilinear coordinate system in a neighbourhood of (\ref{arbcon1}). We suppose that $z_n$ is just
a distance from the surface (with a proper sign, of course) and coordinate lines of $z_1, z_2, \ldots z_{n-1}$ are orthogonal to that of $z_n$.
We propose the following choice of function $f(x)$: it should be equal $z_n$. 
Such smooth function exists in the whole vicinity of any orientable surface. After that we have
$|\overrightarrow\bigtriangledown f|=1$ and $\partial_i n_k=\partial_k n_i$, 
$\sum\limits_{k=1}^nn_k\partial_kn_i=0$ where $n_k=\partial_k f$. 

\begin{theorem}
The Dirac Hamiltonian for our choice of the function f(x) is
${\hat H}^{\cal (D)}=-\frac{\hbar^2}{2}\Delta_{LB}+\frac{\hbar^2}{8}\left(\sum\limits_{\alpha=1}^{n-1}k_{\alpha}\right)^2$.
\end{theorem}
\noindent {\it Proof.} 
The quantum potential is
\begin{multline*}
V_q^{\cal(D)}=\frac{\hbar^2}{4}\sum\limits_{i=1}^n\left(\partial_i-n_i\sum\limits_{k=1}^nn_k\partial_k\right)
\left(\sum\limits_{j=1}^n\partial_jn_in_j\right)-\frac{\hbar^2}{8}(\sum\limits_{j=1}^n\partial_jn_in_j)^2\\
=\frac{\hbar^2}{4}\left(\left(\div(\overrightarrow n)\right)^2+\sum\limits_{i=1}^nn_i\partial_i\cdot\div(\overrightarrow n)-
\sum\limits_{k=1}^nn_k\partial_k\cdot\div(\overrightarrow n)\right)-\\-\frac{\hbar^2}{8}\left(\div(\overrightarrow n)\right)^2=
\frac{\hbar^2}{8}\left(\div(\overrightarrow n)\right)^2=\frac{\hbar^2}{8}\left(\sum\limits_{\alpha=1}^{n-1}k_{\alpha}\right)^2.
\end{multline*}
For spheres it yields the previous result $\frac{\hbar^2(n-1)^2}{8R^2}$.

The kinetic part of the of ${\hat H}^{\cal (D)}$ is obtained by the following lemma:

\begin{lemma}
The Laplace-Beltrami operator on the surface $f(x)=0$ is 
$$\Delta_{LB}={\tilde\Delta}-\left(\frac{\partial}{\partial\overrightarrow n}\right)^2-\div({\overrightarrow n})\cdot\frac{\partial}{\partial \overrightarrow n}.$$
\end{lemma}
\begin{proof}
 In curvilinear coordinates $z_i$
the metric tensor is
$${\tilde g}_{ik}=\left (
\begin{matrix}
g_{\alpha\beta}&0 \\ 
0&1
\end{matrix}
\right).$$ The definition of Laplace operator reads $${\tilde\Delta}=\sum\limits_{i=1}^n\sum\limits_{k=1}^n{\tilde g}^{-1/2}\partial_i{\tilde g}^{1/2}{\tilde g}^{ik}\partial_k
={\partial_n}^2+\left({\tilde g}^{-1/2}\partial_n{\tilde g}^{1/2}\right)\partial_n+\Delta_{LB}$$ with $\Delta_{LB}$ being the Laplace-Beltrami
operator on a surface $z_n=const$. The constraint (\ref{arbcon1}) is $z_n=0$. Let's take another surface, $z_n=\epsilon$:

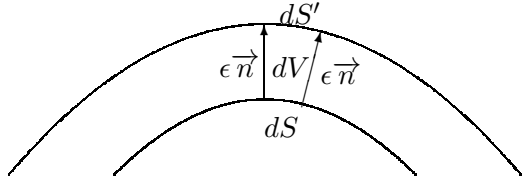
\begin{picture}(100,30)
\qbezier(30,0)(50,20)(70,0)
\qbezier(16,0)(50,40)(84,0)
\put(50,10){\vector(0,1){10}}
\put(55,9){\vector(1,4){2.5}}
\put(50,5){$dS$}
\put(52,20){$dS^{\prime}$}
\put(51,13){$dV$}
\put(44,13){$\epsilon\overrightarrow n$}
\put(57.5,12){$\epsilon\overrightarrow n$}
\end{picture}

\vspace{4ex}
We have $\div({\overrightarrow n})=\frac{dS^{\prime}-dS}{dV}+{\cal O}(\epsilon)=\frac{dS^{\prime}-dS}{\epsilon dS}+{\cal O}(\epsilon)$, hence
$dS^{\prime}=dS(1+\epsilon\div(\overrightarrow n)+{\cal O}(\epsilon^2))$ and ${\tilde g}^{-1/2}\partial_n{\tilde g}^{1/2}=\div(\overrightarrow n)$.
It proves that ${\tilde\Delta}={\partial_n}^2+\div(\overrightarrow n)\cdot\partial_n+\Delta_{LB}$ and
${\hat H}^{\cal (D)}=-\frac{\hbar^2}{2}\Delta_{LB}+\frac{\hbar^2}{8}\left(\sum\limits_{\alpha=1}^{n-1}k_{\alpha}\right)^2$, while
${\hat H}^{\cal (P)}=-\frac{\hbar^2}{2}\Delta_{LB}$ exactly as in Podolsky theory with $V_q^{\cal (P)}=0$.
\end{proof}

In general the unit normal vector ${\overrightarrow n}=\frac{\overrightarrow\bigtriangledown f}
{|{\overrightarrow\bigtriangledown f}|}$ would not be orthogonal to the surfaces $z_n=const\neq 0$ and the result of
the Lemma 5 would not be true. The second normal derivative $\left(\frac{\partial}{\partial\overrightarrow n}\right)^2$ would yield
an additional first order differential term to $\Delta_{LB}$. For a parabola $f(x)=x_2-\frac{k}{2}x_1^2=0$ we have
$n_1=\frac{kx_1}{\sqrt{1+k^2x_1^2}}$ and $n_2=-\frac{1}{\sqrt{1+k^2x_1^2}}$. One can easily see
that at the surface $f(x)=0$
$$\frac{\partial^2}{\partial x_1^2}+\frac{\partial^2}{\partial x_2^2}-\frac{\partial^2}{\partial{\overrightarrow n}^2}-
\div\left(\overrightarrow n\right)\frac{\partial}{\partial\overrightarrow n}=\Delta_{LB}-\frac{k^2x_1}{(1+k^2x_1^2)^2}\frac{\partial}{\partial x_1},$$
where $\Delta_{LB}=\frac{\partial}{\partial\overrightarrow t}$ with $t_1=\frac{1}{\sqrt{1+k^2x_1^2}}$,
$t_2=\frac{kx_1}{\sqrt{1+k^2x_1^2}}$. So, the kinetic part of the Hamiltonian in the Dirac
recipe is also ambiguous.

\section{Some remarks on abelian conversion}
The abelian conversion method \cite{Faddeev,Batalin} consists of introducing new canonical pair of variables
$Q,\ K$ and first class constraints $\sigma_1,\ \sigma_2$: $\{\sigma_1,\sigma_2\}=0$; 
$\sigma_1=\phi_1,\ \sigma_2=\phi_2$ if $Q=0$ and $K=0$. In our case it would be
$\sigma_1=f(x)+K$ and $\sigma_2=\overrightarrow n\cdot\overrightarrow p+Q$. The next step
is to find a new Hamiltonian such that $H_S=H$ if $Q=0$ and $K=0$ and 
$\{H_S,\sigma_1\}=\{H_S,\sigma_2\}=0$. The physical sector is obtained by setting
$\sigma_1=\sigma_2=0$.

For a sphere it yields zero quantum potential, see \cite{KleSha,KlaSha}. In this section
we point out some difficulties in the way of applying this method to arbitrary surfaces.
For spheres the authors of \cite{KleSha,KlaSha} had the result of the form
$H_S=H_S(\sigma_1,\sigma_2,\sum\limits_{i<k}(x_ip_k-x_kp_i)^2)$ and it is not difficult to see
that $$\sum\limits_{i<k}(x_ip_k-x_kp_i)^2=(\sum\limits_ix_i^2)(\sum\limits_ip_i^2
-(\sum\limits_in_ip_i)^2).$$ It allows to get the correct answer because
$\sigma_2^2=(\sum\limits_in_ip_i)^2$ if $Q=0$.

In general case let's try to search for $H_S$ in a form $$H_S=H_S(\sigma_1,\sigma_2,g(x)(\sum\limits_ip_i^2
-(\sum\limits_in_ip_i)^2).$$ The equation for $g(x)$ can be obtained by use of relations
\begin{eqnarray}
\label{relat1}
\{H_S,\sigma_1\}=-\sum_in_i\frac{\partial H_S}{\partial p_i}=0,\\
\label{relat2}
\{H_S,\sigma_2\}=\sum_in_i\frac{\partial H_S}{\partial x_i}-\sum_{i,k}p_k(\partial_k n_i)\frac{\partial H_S}{\partial p_i}=0.
\end{eqnarray}
From (\ref{relat2}) we have
$$\sum_{i,k}p_ip_k(n_j\partial_jg(x)(\delta_{ik}-n_in_k)-2g(x)(\partial_in_k))=0.$$
For spheres it has a non-zero solution because $\partial_in_k\sim\delta_{ik}-n_in_k$. But this is not
true for arbitrary surfaces. Hence the result of \cite{KleSha} can't be
generalized directly.

Moreover, we show that on this way quadratic in momenta $p_i$ physical Hamiltonian
is not possible in general. Let's try to find it in a form
$$H_S=H_S(\sigma_1,\sigma_2,\sum_{i,k}C_{ik}(x)p_ip_k+\sum_iD_ip_i+E(x))$$
with symmetric matrix $C_{ik}$.
Equations (\ref{relat1}) and (\ref{relat2}) yield:
\begin{eqnarray}
\label{C1}
\sum_in_iC_{ik}=0,\ \forall\ k,\\
\label{C2}
\sum_in_i\partial_iC_{lk}-\sum_iC_{il}\partial_in_k-\sum_iC_{ik}\partial_in_l=0,\ \forall\ l,k,
\end{eqnarray}
\begin{eqnarray*}
\sum_in_iD_i=0,\\
\sum_in_i\partial_iD_k-\sum_iD_i\partial_in_k=0,\ \forall\ k,\\
\sum_in_i\partial_iE=0.
\end{eqnarray*}
For $C_{ik}$ we have more equations than variables. This system does not have non-zero
solution in general case. The problem appears even for a parabola, $f(x)=x_2-\frac{k}{2}x_1^2$.
From (\ref{C1}) one has $C_{11}=kx_1C_{12}$ and $C_{22}=\frac{C_{12}}{kx_1}$. After that
(\ref{C2}) turns to yield three different equations for one function $C_{12}(x)$. This system
is not solvable.

One could consider a coordinate system $z_1,z_2$ from section 3 with $n=2$. Then
 we have $C_{22}=C_{12}=0$ and $\partial_{z_2}C_{11}=0$. This system is solvable,
of course, but the coodinates $z$ are not Cartesian.

Let's consider Cartesian coodinates and function $f(x)=z_n$. In this case the
unit normal equals $n_i=\partial_if$ and has some additional properties:
$\partial_in_k=\partial_kn_i$; $\sum\limits_{i=1}^nn_i\partial_in_k=0$.
With these properties equations (\ref{C1})-(\ref{C2}) are solvable. Indeed, we have
$$C_{\alpha n}=-\frac{1}{n_n}\sum\limits_{\beta=1}^{n-1}n_{\beta}C_{\alpha\beta}, \quad
C_{nn}=\frac{1}{n_n^2}\sum\limits_{\alpha=1}^{n-1}\sum\limits_{\beta=1}^{n-1}
n_{\alpha}n_{\beta}C_{\alpha\beta}$$
from (\ref{C1}) and analogous relations for the normal
$$\partial_nn_{\alpha}=\partial_{\alpha}n_n=-\frac{1}{n_n}\sum\limits_{\beta=1}^{n-1}
n_{\beta}\partial_{\beta}n_{\alpha},\quad \partial_nn_n=\frac{1}{n_n^2}
\sum\limits_{\alpha=1}^{n-1}\sum\limits_{\beta=1}^{n-1}n_{\alpha}n_{\beta}\partial_{\alpha}n_{\beta}.$$
After that for $C_{\alpha,\beta}$ equation (\ref{C2}) yields
\begin{multline}
\label{I}
\sum_{i=1}^nn_i\partial_iC_{\alpha\beta}-\sum_{\zeta=1}^{n-1}\sum_{\gamma=1}^{n-1}C_{\alpha\zeta}\partial_{\gamma}n_{\beta}
\left(\delta_{\zeta\gamma}+\frac{n_{\zeta}n_{\gamma}}{n_n^2}\right)-\\-
\sum_{\zeta=1}^{n-1}\sum_{\gamma=1}^{n-1}C_{\beta\zeta}\partial_{\gamma}n_{\alpha}
\left(\delta_{\zeta\gamma}+\frac{n_{\zeta}n_{\gamma}}{n_n^2}\right)=0
\end{multline} 
To this moment everything is solvable
(provided that we made a good choice of direction of $n$-th axis). And it's not difficult to see
that remaining equations in (\ref{C2}) take the form
\begin{multline*}
-\frac{1}{n_n}\sum_{\alpha=1}^{n-1}n_{\alpha}\left[
\sum_{i=1}^nn_i\partial_iC_{\alpha\beta}-\sum_{\zeta=1}^{n-1}\sum_{\gamma=1}^{n-1}C_{\alpha\zeta}\partial_{\gamma}n_{\beta}
\left(\delta_{\zeta\gamma}+\frac{n_{\zeta}n_{\gamma}}{n_n^2}\right)-\right.\\-\left.
\sum_{\zeta=1}^{n-1}\sum_{\gamma=1}^{n-1}C_{\beta\zeta}\partial_{\gamma}n_{\alpha}
\left(\delta_{\zeta\gamma}+\frac{n_{\zeta}n_{\gamma}}{n_n^2}\right)\right]=0,
\end{multline*}
\begin{multline*}
\frac{1}{n_n^2}\sum_{\alpha=1}^{n-1}\sum_{\beta=1}^{n-1}n_{\alpha}n_{\beta}\left[
\sum_{i=1}^nn_i\partial_iC_{\alpha\beta}-\right.\\-\left.
2\sum_{\zeta=1}^{n-1}\sum_{\gamma=1}^{n-1}C_{\alpha\zeta}\partial_{\gamma}n_{\beta}
\left(\delta_{\zeta\gamma}+\frac{n_{\zeta}n_{\gamma}}{n_n^2}\right)\right]=0
\end{multline*}
and follow directly from (\ref{I}).

So, quadratic in momenta Hamiltonian is possible in Cartesian coordinates if
one admits the special definition of function $f(x)$. But even after that this method
can't yield the Podolsky theory for arbitrary surface, because it would mean that the quantum
physical Hamiltonian is $-\frac{\hbar^2}{2}\Delta_{LB}$ on the surface. And it follows
from Lemma 5 that for it to take place with momenta ${\hat p_i}=-i\hbar\frac{\partial}{\partial x_i}$,
we should have in the classical limit a Hamiltonian with quadratic in
momenta term proportional to $\sum\limits_ip_i^2
-(\sum\limits_in_ip_i)^2$. But generally it's not the case.

\section{Conclusion}
Quantum theory contains more information than classical one.
That's why it is not possible to find a unique quantization for a given classical
theory. In the quantum world it is no longer enough to say that some particle moves
in a certain curved space. One should know the nature of this motion. If it's just some
potential force which makes the particle to stay at the curved surface, one should
use the thin layer quantization method \cite{Costa,Jensen,me}. But if there is no outer space (apart
from our formalism) the result should not depend on any extrinsic properties of
the physical space, so that Podolsky theory seems to be the preferable one.
We reproduced this theory by our modification of Dirac quantization procedure.
Still there may be some physical systems to which the original Dirac
quantization should be applied. The Dirac recipe turned out to be ambiguous, but
we proposed a natural way to overcome this ambiguity.


\begin{thebibliography}{99}
\bibitem{Batalin} 
Batalin I.A., Fradkin E.S., {\it Nucl. Phys. B}\quad {\bf 279}, 514 (1987).
\bibitem{Brat1} 
Bratchikov A.V., {\it Lett. Math. Phys.}\quad {\bf 61}, 107 (2002); {\it preprint} hep-th/0204019.
\bibitem{Brat2} 
Bratchikov A.V., {\it preprint} hep-th/0312240.
\bibitem{Costa} 
da Costa R.C.T., {\it Phys. Rev. A} {\bf 23}, 1982 (1981).
\bibitem{Dirac} 
Dirac P.A.M., {\it Lectures on quantum mechanics} (Yeshiva University, N.Y., 1964).
\bibitem{Faddeev} 
Faddeev L.D., Shatashvili S.L., {\it Phys. Lett. B}\quad {\bf 167}, 255 (1986).
\bibitem{me}
Golovnev A.V., {\it preprint} quant-ph/0508111.
\bibitem{Jensen}
Jensen H., Koppe H., {\it Ann. Phys.} {\bf 63}, 586 (1971).
\bibitem{KlaSha} 
Klauder J.R., Shabanov S.V., {\it Nucl. Phys. B}\quad {\bf 511}, 713 (1998); {\it preprint} hep-th/9702102.
\bibitem{KleSha} 
Kleinert H., Shabanov S.V., {\it Phys. Lett. A}\quad {\bf 232}, 327 (1997); {\it preprint} quant-ph/9702006.
\bibitem{Reyman} 
Reyman A.G., Semenov-Tian-Shansky M.A., {\it Integriruemie sistemi} (2-nd ed., Moskva{\&}Izhevsk, 2003)
({\it in Russian}). English translation of a very short version of this work is available in V.~ Arnold, S.P. Novikov editors, {\it Dynamical systems,
VII}, volume 16 of {\it Encyclopaedia of Mathematical Sciences} (Springer-Verlag, Berlin, 1994).
\bibitem{Podolsky} 
Podolsky B., {\it Phys. Rev.} {\bf 32}, 812 (1928).
\end{thebibliography}
\end{document}